Stephane Daul
Institut de Physique Theorique
University of Fribourg, Perolles
1700 Fribourg
Switzerland
                                               Fribourg,  19 March 1996

Dear Editor,

Please find enclosed our manuscript "Low density ferromagnetism in the
Hubbard model" In our opinion it deals with a timely subject, and we
believe it sheds some new light on the old problem of band ferromagnetism.
Therefore we are submitting our manuscript to Physical Review Letters.
The figures are being sent by Fax. The original figures and the Copyright
transfer form will be sent by ordinary mail.

Sincerly yours,
                         
+++++++++++++++++++++++++++++++++++++++++++++++++++++++++++++++++++++++++++

\documentstyle[prl,aps]{revtex}
\begin{document}

\tolerance=10000

\twocolumn[\hsize\textwidth\columnwidth\hsize
     \csname @twocolumnfalse\endcsname

\title{Low density ferromagnetism in the Hubbard model}
\author{P. Pieri\\
{\small {\em Dipartimento di Fisica, INFM and INFN, 
Universit\`a di Bologna,}}\\
{\rule[-3mm]{0mm}{5mm}{\small {\em Via Irnerio 46, I-40126, Bologna, Italy.}}}
\\
S. Daul, D. Baeriswyl,  M. Dzierzawa\\
{\rule[-3mm]{0mm}{5mm}{\small {\em Institut de Physique Th\'{e}orique, 
Universit\'{e}, P\'erolles, 
CH-1700 Fribourg, Switzerland.}}}\\
P. Fazekas\\
{\small {\em Research Institute for Solid State Physics, Budapest 114, POB 49,
H-1525 Hungary.}}}
\maketitle
\begin{abstract}
A single-band Hubbard model with nearest and next-nearest neighbour hopping 
is studied for $d=1$, 2, 3, using both analytical and numerical techniques. 
In one dimension, saturated ferromagnetism is found above a critical value 
of $U$ for a band structure with two minima and for small and intermediate 
densities. This is an extension of a scenario recently proposed by 
M\"uller--Hartmann. For three dimensions and non-pathological band structures, 
it is proven that such a scenario does not work.
\end{abstract}

\vskip2pc]
\newpage
 
The Hubbard Hamiltonian \cite{lie93}, which has recently attracted so much 
interest as a 
model for describing high-T$_c$ superconductors, was investigated already in 
the sixties to face the problem of ferromagnetism in itinerant
electron systems \cite{hub63,kan63}. Indeed, at a mean-field level, the 
Hubbard model seems to be a good starting point, as the Stoner criterion
predicts a ferromagnetic ground state in a wide range of parameters. 
However, the inclusion of correlation effects 
makes the conditions for the appearance of ferromagnetism more 
stringent \cite{hub63}.
The existence of large-spin ground states
in Hubbard type models has been rigorously established in three 
different but rather peculiar situations: 
Nagaoka ferromagnetism for infinite $U$ and one hole in a half-filled 
band \cite{nag66}, 
Lieb ferrimagnetism for half-filled bipartite lattices with 
sublattices containing a different number of sites \cite{lie89}, and the flat
(or nearly-flat) band ferromagnetism of Mielke \cite{mie93} and 
Tasaki \cite{tas,tas95}.

In a recent paper M\"{u}ller--Hartmann proposed an alternative route to 
ferromagnetism \cite{mh95}. 
He considered the one-dimensional Hubbard model with both nearest- and 
next-nearest-neighbour hopping ($t_1$--$t_2$ Hubbard model).
For infinite $U$ and a band dispersion with two degenerate minima he found 
saturated ferromagnetism for small electron densities. Very recently Penc and 
coworkers have studied a generalized version of Tasaki's model in $1d$ and 
found ferromagnetism in a large region of parameters \cite{pen}. Their model 
reduces to that of M\"uller--Hartmann in a certain limit to which,
however, their considerations do not apply.

In this paper we present numerical and variational results for the $1d$ 
$t_1$--$t_2$ Hubbard model, which do confirm the existence of 
M\"uller--Hartmann ferromagnetism. Indeed we find that this phase is not
confined to infinite $U$ and vanishingly small densities but occupies a
substantial portion of parameter space. It is then natural to ask whether 
this route to ferromagnetism works also for higher dimensions. We will show 
that in dimensions $d>2$ the answer is negative. For low enough densities 
(and a non-pathological band structure), even in the presence of several 
degenerate minima any state with finite magnetization is unstable. For the 
``marginal'' dimension $d=2$ we cannot exclude ferromagnetism but rather 
restrict strongly the region of its stability. In fact, the existence of many 
degenerate minima is not enough to stabilize the
fully polarized state: what really matters is a high density of states at the
bottom of the band.
   
We first consider the one-dimensional case. For the simple Hubbard model with 
only nearest-neighbour hopping, ferromagnetism has been discarded a long time 
ago by Lieb and Mattis, who showed that the ground state is a singlet 
\cite{lm}. For the $t_1$--$t_2$ Hubbard chain
\begin{equation}
H = -\sum_{i\sigma}(t_1 c^{\dagger}_{i\sigma} c_{i+1\sigma}
+t_2 c^{\dagger}_{i\sigma} c_{i+2\sigma} + {\mbox{\rm h.c.}})
+ U\sum_i n_{i\uparrow}n_{i\downarrow}   
\end{equation}
the Lieb--Mattis theorem is not applicable since when $t_2\neq 0$ the 
particles can pass each other and so cannot be ordered, which would be 
essential for the proof of the theorem.  Actually, Mattis and Pe\~{n}a have 
shown in \cite{mat74} that, when $t_1>0$ and $t_2<0$, the model shows Nagaoka 
ferromagnetism for one hole in a half-filled band. Here we are interested 
in small and intermediate densities.

We have performed numerical diagonalizations using the Lanczos method for 
chains up to $L=18$ sites, for even electron numbers $N\le L$ ($n=N/L$ is the 
electron density). The total spin of the 
ground state is determined both by measuring the spin--spin correlation 
function at $q=0$ and by calculating the lowest energy states for each $S_z$ 
subspace. The energy of the fully 
polarized state is independent of $U$ and calculated analytically.
When $t_2>-t_1/4$ (we always suppose $t_1 > 0$) the band has only one minimum 
at $k=0$ and we do not find
ferromagnetism at low density: for every value of $U$ the ground state is a 
singlet. From now on we consider only the case $t_2<-t_1/4$, when the band 
developes two degenerate minima at momenta
$\pm k_0$, with $k_0=\arccos(t_1/4t_2)$. It is in this region that we should 
find, according to M\"{u}ller--Hartmann, ferromagnetism at low density and 
infinite $U$. We note that the lowest density that can be studied on a finite
chain is $2/L$. For this case of two electrons, one can 
easily show that the ground state is a 
triplet. However, this result can be also regarded as an example of Mielke's 
flat band ferromagnetism. Interestingly,
for large enough $U$, we do find ferromagnetism also for more than two 
electrons, namely for a range of densities up to a critical
value $n_c$. This is no longer flat band ferromagnetism but rather
confirms the validity of M\"{u}ller--Hartmann's construction beyond
the limit $n\to 0$ and $U=\infty$.

In our numerical diagonalizations we have never found partially
polarized ground states: the phase transition is always between a singlet 
and the fully polarized state. Fig.1 shows the behaviour of $n_c$ as a 
function of $t_2/t_1$. Data from chains of different lengths are plotted 
in the following way: if for instance with $N=4$ electrons the ground state is 
ferromagnetic, while with $N=6$ it is not, we draw an error bar between $4/L$ 
and $6/L$. The solid curve represents the density at which the 
number of Fermi points for the fully polarized state changes from
four to two. The curve seems to fit the data rather well and shortly, we are 
going to present an analytical argument that this is the likely criterion for 
the phase boundary\footnote{At least for not too large $|t_2|$. In fact, for 
$\vert{t_2}\vert/t_1\to\infty$, the system decouples into two simple Hubbard 
chains.}. In any case, one can say that for densities above this curve 
M\"{u}ller--Hartmann's construction is no longer applicable
since it relies on the existence of two disconnected pockets. In summary, our 
results indicate that this type of ferromagnetism is 
not restricted to very low densities, but it survives as long as the 
(fully polarized) Fermi sea has four Fermi points. 

We turn now to the question which value of $U$ is required to sustain 
ferromagnetism at a given density $n$. Since our numerical results did not 
show evidence for a partially polarized ground state, we may  
examine the stability of the fully polarized state $\vert F\rangle$ with 
respect to a single spin flip. We use the standard single-spin-flip trial state 
\cite{sha90}
\begin{equation}
\vert\psi\rangle=\prod_i(1-\eta n_{i\uparrow}n_{i\downarrow})
c^{\dagger}_{k_0\downarrow}c_{k_F\uparrow}\vert F\rangle .  
\label{eq:vas}
\end{equation}
An up-spin electron is removed from one of the Fermi points $k_F$ (with energy 
$\epsilon_F$) and a  down-spin electron is put to the bottom of one of the 
degenerate minima $\epsilon_0$ at wave vector $k_0$. This yields a 
single-particle energy gain of $\Delta=\epsilon_F-\epsilon_0$.
The change of the total energy is
\begin{eqnarray}
\Delta E&=&-\Delta + U n \frac{(1-\eta)^2}{1+(\eta^2-2\eta)n}+
\frac{\eta^2}{1+(\eta^2-2\eta)n} \nonumber \\[2mm]
& & \cdot\left[\epsilon_0(n^2-n)
-\frac{E_0}{L}+\frac{E_{1}^2 \cos k_0}{2t_1L^2} 
+\frac{E_{2}^2\cos 2k_0 }{2t_2L^2}\right]
\label{eq:ene}
\end{eqnarray}
where $E_0=E_1+E_2$ is the energy of the fully polarized state, while 
$E_1$ and $E_2$ are the contributions from
the nearest-neighbour and next-nearest-neighbour hopping terms, respectively.
$\Delta E$ has to be minimized with respect to the variational parameter 
$\eta$.

The polarized state becomes unstable when the kinetic energy gain outweighs 
the increased interaction energy.
Fig.2. shows, for the special value $t_2=-t_1$ the variational phase 
boundary $U_{\rm var}(n)$  (continuous curve) and the critical $U_c$ 
determined by the numerical diagonalization (dots). We found similar 
results for other values of $t_2$ ($<-t_1/4$). We notice that $U_{\rm var}$ 
is a lower bound for the exact $U_c$. For small density $n$, $U_c$ decreases 
with decreasing $n$, apparently tending to zero as $n\to 0$. The cusp of 
$U_{\rm var}(n)$ occurs at precisely the critical density 
we have defined above and indicates that something drastic happens there. 
While for $n>n_c$ we see a sharply  
rising $U_{\rm var}(n)$, our numerical data indicate that ferromagnetism is 
completely suppressed in this region. Since the variational result provides a 
quickly increasing lower bound for $U_c$, we take it as an indication 
that, in fact, $U_c=\infty$ for $n>n_c$. In this sense, the variational 
approach supports the phase diagram shown in Fig. 1.

We investigate now the stability of low-density ferromagnetism for $d\ge 2$. 
We consider the Hubbard Hamiltonian with arbitrary non-diagonal hopping terms 
$t_{ij}$ and vanishing diagonal terms, $t_{ii}=0$. However, we confine
ourselves to non-pathological band structures, meaning quadratic dispersion
about the minima. Since the most favourable 
situation for a fully polarized state is $U=\infty$, we use the variational 
ansatz (\ref{eq:vas}) for $\eta=1$. A straightforward calculation gives
\begin{equation}
\Delta E=-2n\epsilon_0-\Delta + O(n^{1+2/d})
\label{eq:var}
\end{equation} 
where $\Delta=\epsilon_F-\epsilon_0\sim n^{2/d}$. (In $1d$, the leading 
correction would be $O(n^2)$.) We note that $\epsilon_0$, the lowest 
eigenvalue of the hopping matrix, is 
necessarily negative, as the sum of the eigenvalues is equal to 
$\sum_i t_{ii}=0$. The first term in Eq. (\ref{eq:var}) is therefore positive, 
favouring ferromagnetism, while the second term destabilizes the fully 
polarized state. 

We examine now the viability of the low-density route for dimensions $d>1$. 
For $d\ge 3$ the second term in (\ref{eq:var}) is the leading term at low 
density and the polarized state is always unstable for sufficiently low 
densities. (For $d=1$ the first term is dominating and the present argument 
does not predict a destabilization of the fully polarized state.) For 
$d=2$ the two terms are of the same order and must be
compared more carefully. In this case, to leading order in $n$, 
$\Delta=n/\rho(\epsilon_0)$, where $\rho(\epsilon_0)$ is the density of 
states at the bottom of the band. Inserting this expression into Eq. 
(\ref{eq:var}) and using the 
stability criterion $\Delta E > 0$, we obtain the following necessary  
condition for the stability of the fully polarized state,
\begin{equation}
2\rho(\epsilon_0)|\epsilon_0|>1.
\label{eq:stab}
\end{equation}
To be specific, let us consider the $t_1$--$t_2$ Hubbard model on a square 
lattice. When $t_2<-t_1/2$ the band has degenerate minima at the Brillouin 
zone boundary. It follows from Eq. (\ref{eq:stab}) that the polarized state 
cannot be the ground state outside the region $-0.20t_1>t_2>-0.65t_1$.
Hence for $t_2<-0.65t_1$ the polarized state is unstable even though there 
are two minima\footnote{The $t$-matrix expansion has been also
used to study the possibility of low-density ferromagnetism
in the $3d$ and $2d$ Hubbard models \cite{mat,rud85}. Let us remark however that
this expansion, even if physically justifiable, is nevertheless uncontrolled.
Moreover, the role of several equivalent minima was not discussed. For the
case of a single minimum, our results are in agreement with those of 
\cite{mat,rud85}.}. 

Having excluded low-density saturated ferromagnetism for $d>2$, we may wonder 
whether we can also say something about partially polarized states. 
Unfortunately, energy eigenstates with $S<S_{max}$ are not exactly known, so 
that we have to find a different way for judging the stability
of ferromagnetic order. An essential ingredient for the argument is the 
monotonous increase of the lowest energy eigenvalue as a function of $U$,  
within any given subspace. This holds for the lowest singlet energy $E_s(U)$ 
as well as for the lowest energy $E_{pp}(U)$ within a subspace with a given 
partial polarization, as depicted in Fig. 3. An additional important fact is 
that the Hartree--Fock state gives an upper bound for the the lowest singlet 
state $E_{HF}(U)>E_s(U)$. Therefore the energies for the partially polarized 
state and the lowest singlet state cannot cross at a smaller $U$ than the 
value defined by $E_{HF}(U)=E_{pp}(0)$ (see Fig. 3). But this value of $U$ 
becomes arbitrarily large as the density goes to zero. In fact, for $d>2$ it 
diverges as $n^{2/d-1}$.

We should comment on the relationship between our description of the
low-density route to ferromagnetism and other scenarios of ferromagnetism in 
Hubbard models. The existence of the single-hole Nagaoka state is well-known 
\cite{mat74} but it seems clear that the fully polarized state of the $1d$ 
$t_1$--$t_2$ model at low to intermediate densities is in its physical 
nature different, and in the phase diagram disconnected, from the Nagaoka 
state. In the special case $n=2/L$, our ferromagnetism is indistinguishable 
from Mielke's \cite{mie93} but we find that the ordered state extends up to a 
($t_2/t_1$-dependent) critical $n_c$, and for finite concentrations $n$ we 
certainly do not have flat-band ferromagnetism.    

We have already mentioned that the generalization of Tasaki's model 
\cite{tas95} by Penc et al. \cite{pen} includes M\"uller--Hartmann's model 
\cite{mh95} as a special case. However, the understanding of ferromagnetism 
in \cite{pen} is based on the perturbative treatment of ring exchange 
processes, and this scheme breaks down for the model considered by us. Our 
finding that a decisive role is played by the existence of four Fermi points, 
seems to have no counterpart in their reasoning. The possible relationship 
between the ferromagnetic phase discussed in \cite{pen} and the 
one described by us, remains to be clarified.
 
It is worthwhile to point out that the existence of several degenerate minima 
does not play an essential role in our reasoning for $d\ge 2$. The phenomenon 
of valley-degeneracy-assisted 
ferromagnetism is apparently restricted to $1d$ systems.
 
To conclude, we have found an extended region of fully polarized ferromagnetic
states in the $U$--$n$ plane for the $1d$ $t_1$--$t_2$ model. The available 
evidence indicates that ferromagnetism exists whenever the Fermi sea consists
of two disconnected parts. The ferromagnetism described by us is not a 
flat-band phenomenon, and it does not seem to be connected with the Nagaoka 
state, either. It is most unfortunate that the mechanism giving rise to the 
spin-polarized state in $1d$ does not work for non-pathological situations in 
$3d$  and has, at best, a rather restricted  range of applicability in $2d$.  

P.P. and P.F. are grateful for the warm hospitality they enjoyed at the 
University of Fribourg, where most of this work was 
done. Financial support by the Swiss National Foundation through the grant 
No. 20-40672.94 is gratefully acknowledged. In Hungary, P.F. was  supported 
by the Hungarian National Research Foundation Grant OTKA T-014201. We are  
grateful to D.C. Mattis, E. M\"uller--Hartmann, K.Penc, H. Shiba and W. von 
der Linden for useful discussions and inspiring correspondence.

\leftline{\bf Figure Captions}

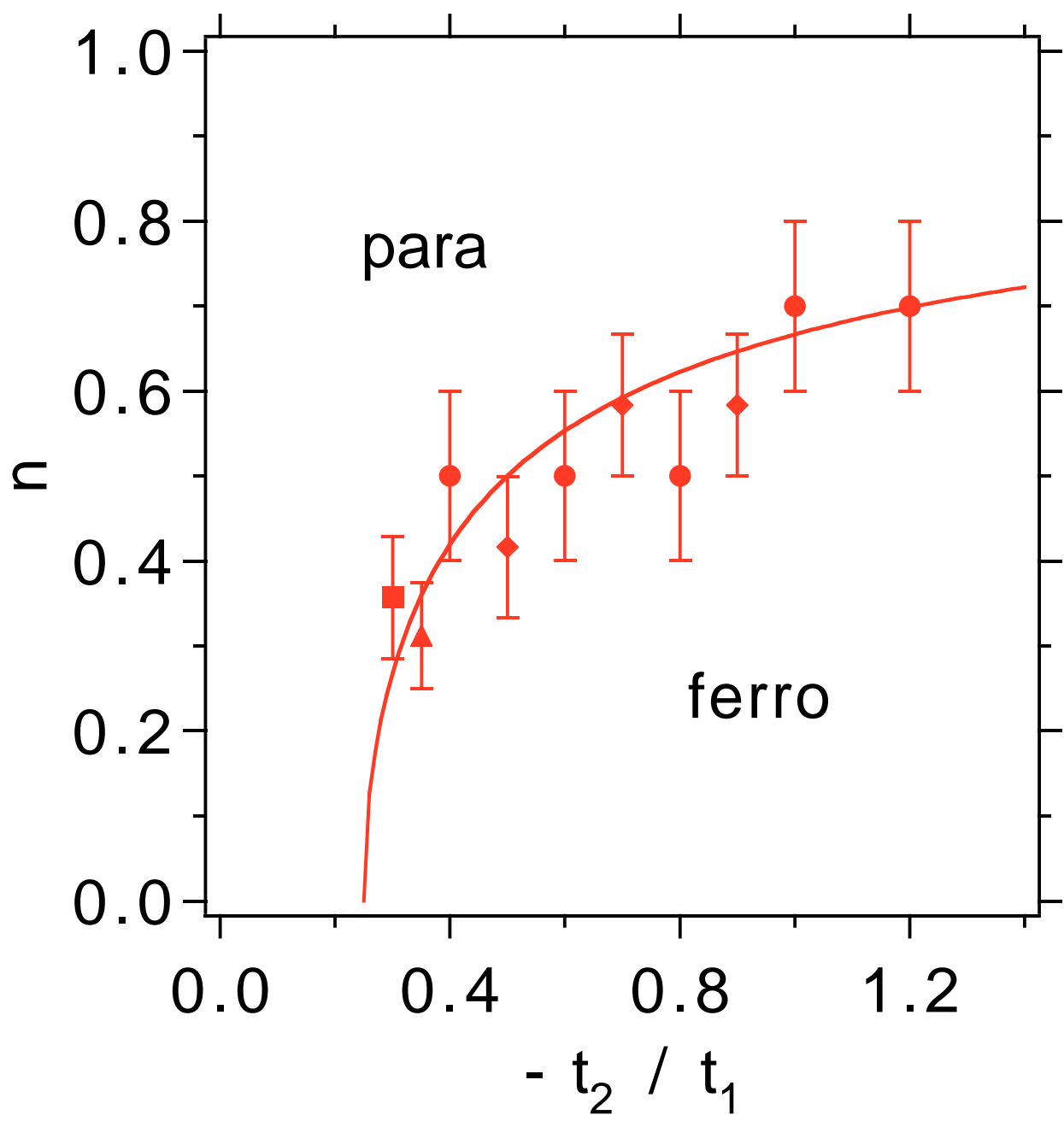
Fig. 1. Phase diagram of the $t_1$--$t_2$ Hubbard chain in the $t_2$--$n$ 
plane, as obtained by exact diagonalization for finite lengths $L$. Circles: 
$L=10$, diamonds: $L=12$, squares: $L=14$, triangles: $L=16$. The full line 
shows the density where the fully polarized Fermi sea splits into two pockets.

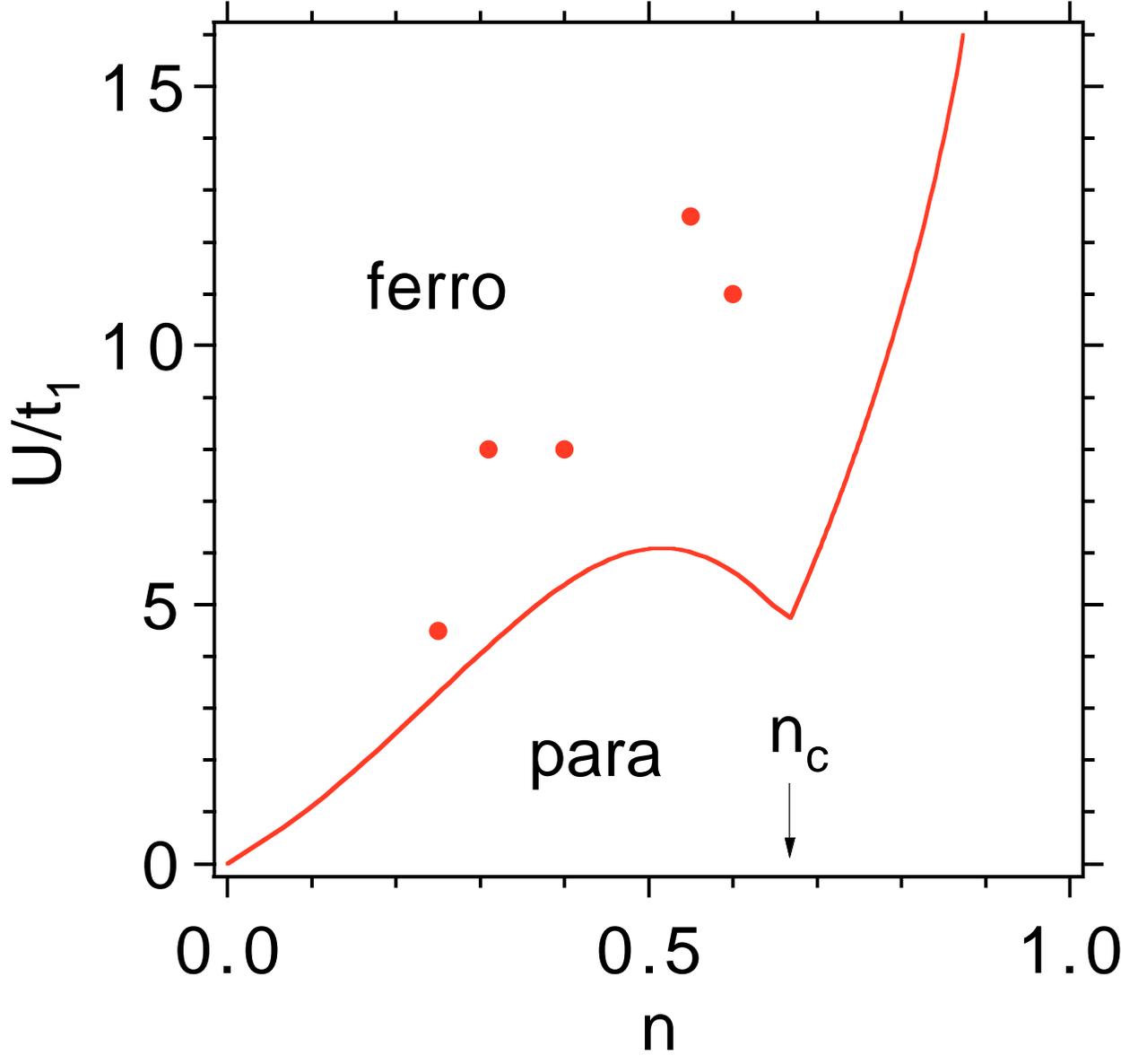
Fig. 2. Phase diagram of the $t_1$--$t_2$ Hubbard chain in the $n$--$U$ plane, 
for $t_2=-t_1$. Circles show numerical results, while the full line gives a 
variational lower bound. The location of the cusp ($n_c=2/3$) corresponds to 
a point on the full line in Fig. 1.

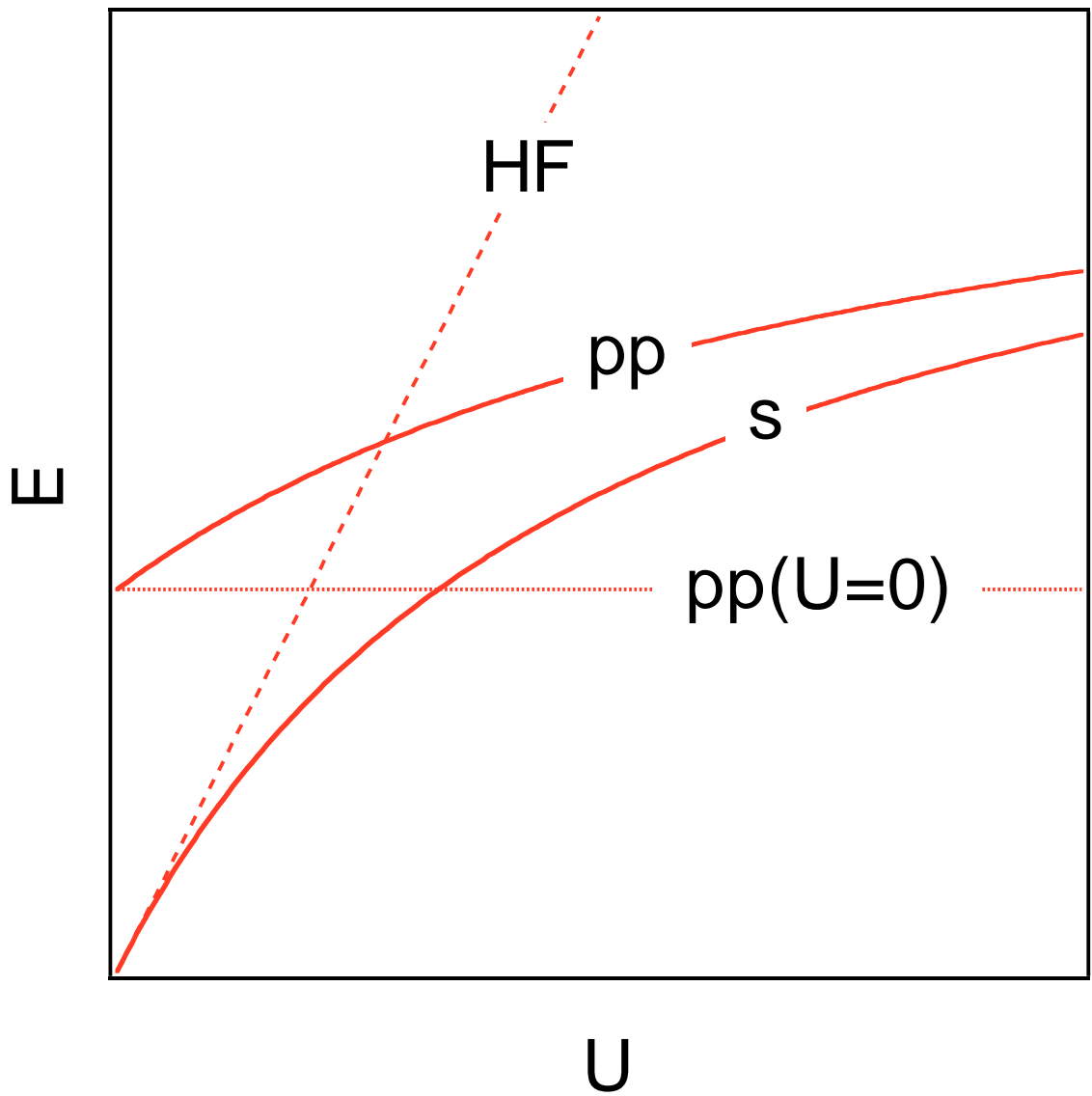
Fig. 3. Schematic view of the $U$-dependence of the lowest energies for
subspaces with different values of the total spin (s: singlet, pp: partially 
polarized). The dashed line shows the non-magnetic Hartree--Fock energy (HF).

\end{document}